\documentclass[fleqn,usenatbib]{mnras}

\usepackage{newtxtext,newtxmath}

\usepackage[T1]{fontenc}

\DeclareRobustCommand{\VAN}[3]{#2}
\let\VANthebibliography\thebibliography
\def\thebibliography{\DeclareRobustCommand{\VAN}[3]{##3}\VANthebibliography}

\usepackage{graphicx}	
\usepackage{amsmath}	
\usepackage{deluxetable} 
\usepackage{adjustbox}
\usepackage{graphicx}
\usepackage{placeins}
\usepackage{float}
\usepackage{tabularx}
\usepackage{threeparttable}
\usepackage{aecompl}

\title[Substellar Components in three sdB Binaries]{Eclipse Timing Modeling of Three Post-Common Envelope Binaries: Hybrid Solutions}

\author[Mai \& Mutel]{
Xinyu Mai,$^{1}$\thanks{E-mail: xinyu-mai@ku.edu}
Robert L. Mutel$^{1}$\thanks{E-mail: robert-mutel@uiowa.edu}\\
$^{1}$Department of Physics and Astronomy, University of Iowa, Iowa City, IA 52242, USA\\
}

\date{Accepted XXX. Received YYY; in original form ZZZ}

\pubyear{2021}

\begin{document}
\label{firstpage}
\pagerange{\pageref{firstpage}--\pageref{lastpage}}
\maketitle

\def\bsq#1{\lq{#1}\rq}
\def \etal{et al.}
\def \eg{e.g.,\ }
\def \HWVir{HW\ Vir\ }
\def \MJ{ M\textsubscript{J}\ }
\def \Mjup{ M\textsubscript{Jup}\ }

\begin{abstract}
We report 90 new observations of three post common envelope binaries at primary eclipse spanning between December 2018 to February 2022. We combine recent primary eclipse timing observations with previously published values to search for substellar circumbinary components consistent with timing variations from a linear ephemeris. We used a least-squares minimization fitting algorithm weighted by a Hill orbit stability function, followed by Bayesian inference, to determine best-fit orbital parameters and associated uncertainties. For HS2231+2441, we find that the timing data are consistent with a constant period and that there is no evidence to suggest orbiting components. For HS0705+6700, we find both one and two-component solutions stable for at least 10 Myr. For HW~Vir, we find three and four-component solutions that fit the timing data reasonably well, but are unstable on short timescales, and therefore highly improbable. Conversely, solutions calculated using a Bayesian orbit stability prior result in a poor fit. The stable solutions significantly deviate from the ensemble timing data in both systems. We speculate that the observed timing variations for these systems, and very possibly other sdB binaries, may result from a combination of substellar component perturbations and an Applegate-Lanza mechanism.
\end{abstract}




\section{Introduction}

Since the discovery of the first extra-solar planets nearly 30 years ago \citep{Wolszczan:1992,Mayor:1995}, more than 4,000 exoplanetary systems have been discovered, mostly orbiting late-type single stars \citep{Martin:2018}. However, more than 100 planet-hosting multiple-star systems have also been found \citep{Schwarz:2016}, perhaps not surprisingly since a substantial fraction of all stars are in multiple systems \citep{Tokovinin:2014}.

An intriguing subset of these planet-hosting binaries are evolved short-period binaries. In these systems, the more massive star has evolved off the main sequence and lost nearly all of its hydrogen envelope during the red-giant phase, filling its Roche lobe and generating unstable mass transfer to the secondary star \citep{Iben:1993}. Since the timescale for mass transfer is much shorter than the thermal timescale of the secondary, a common envelope forms surrounding both components. The orbiting stars experience drag forces, transferring angular momentum to the envelope and causing it to expand and eventually be expelled \citep{Webbink:1984,Zorotovic:2010,Ivanova:2013}. This leaves a hot B-star or white dwarf with a low mass main-sequence companion separated by a few solar radii, and consequent orbital period of a few hours \citep{Nebot:2011, Heber:2009}. These systems are referred to  as post common-envelope binaries (PCEB). 

Eclipsing PCEBs provide a sensitive probe of potential substellar companions, since the eclipses are frequent and deep, and can be timed with high accuracy. This allows one to infer the presence of orbiting substellar companions by modeling the observed eclipse timing variations (ETV's) caused by the reflex motion of the binary's barycenter due to the substellar companion[s].  

To date, more than a dozen eclipsing PCEBs have been observed to exhibit large, occasionally erratic changes in their orbital period \citep{Heber:2016}. Suppose circumbinary substellar components cause these timing variations. In that case, they must have either survived the common envelope phase or formed from the common envelope debris and therefore be much younger than the parent stars. This ‘second generation’ scenario is considered more likely for PCEBs by some authors \citep{Zorotovic:2013, Schleicher:2014, Schleicher:2015}, but this hypothesis is still controversial. For example, \cite{Bear:2014} argue that too large a fraction of the common envelope mass would have to form into substellar companions for this to be viable.

Detailed modeling of eclipse timing variations (ETV) has resulted in the claimed detection of multiple circumbinary companions, either brown dwarfs or massive planets, in at least ten PCEBs \citep{Marsh:2018}. However, these claims are not universally accepted: Several of the proposed planetary orbits have been shown to be dynamically unstable over timescales much shorter than would be consistent with even the ‘second generation’ planetary formation hypothesis \citep{Horner:2012, Horner:2014, Wittenmyer:2013}. 

Alternative explanations for observed eclipse timing variation in binary stars have been considered, including apsidal precession \citep{Shakura:1985, Parsons:2013} and changes in the mass quadruple moment of the secondary star caused by magnetic activity, a.k.a. the Applegate mechanism \citep{Applegate:1987, Applegate:1992,Lanza:1998,Lanza:2006,Beuermann:2012,Lanza:2020}. Apsidal precession, which results from a small non-zero eccentricity of the binary orbit, appears unlikely, since it would require eccentricities of order $e\sim0.01$ to explain observed PCEB eclipse timing variations ($\sim$100 sec). These could be ruled out by observing primary to secondary eclipse timing asymmetries, which would be of the same order as the observed ETVs, and would have a periodic time shift of the secondary eclipse with respect to phase 0.5. In one case (HW Vir), \cite{Baran:2018} reported no significant deviation from phase 0.5 for secondary eclipse times observed over 70 days. Furthermore, ETV’s from PCEB binaries are typically aperiodic, whereas apsidal precession would result in smooth timing variations over an orbital period. 

The Applegate mechanism has been considered to explain PCEB ETV’s by several authors (e.g. \cite{Qian:2008,Lee:2009, Wittenmyer:2012,Nasiroglu:2017}. However, detailed stellar models of the angular momentum exchange between a finite shell and the core of the secondary star show that the required energy to produce the quadrupole moment variations is too high in most PCEB systems, including all three of those described here \citep{Volschow:2016,Navarrete:2018, Navarrete:2020}. However, recently \citep{Lanza:2020} has proposed a new model based on the angular momentum exchange between the spin of the magnetically active secondary and the binary orbital motion. This mechanism may overcome the energy limitations of the original Applegate mechanism, and could contribute to the observed eclipse timing variations.

In this paper, we report new primary eclipse timing observations of three PCEB systems, combining them with previously published observations to solve for possible substellar companions that can account for the observed timing variations.  We also analyze the orbital stability of the proposed systems using numerical simulation of orbital dynamics and testing for chaotic behavior  over long timescales. 

This procedure could be considered fraught and even unproductive, since it is typically degenerate (fitted model parameters are typically highly correlated) and almost all previously proposed planetary models have failed to correctly predict subsequent  ETV observations. Indeed, as \cite{Marsh:2018} recently pointed out, there is no {\it independent} evidence for any substellar companion deduced from ETV variations in PCEBs. Nevertheless, the ETV variations are clearly real and all alternate explanations have so far failed with compelling physical arguments. Hence, both Occam's razor and the scientific method encourage us to assume the most plausible hypothesis, find physical models that are in agreement with the observations, and most importantly, provide predictions that future observations can test.

\subsection{HW Virginis}
\HWVir (V = 10.7, P = 2.8 hr) is the prototype of the eponymous \HWVir class of eclipsing subdwarf B stars with low-mass dwarf companions. \cite{Kilkenny:1994} first discovered variations in the orbital period and reported a period decrease over nine years. Several authors have carried out subsequent ETV studies of HW Vir \citep{Cakirli:1999,Kiss:2000, Kilkenny:2000, Kilkenny:2003, Ibanoglu:2004, Lee:2009, Qian:2010, Beuermann:2012, Skelton:2012,Lohr:2014, Baran:2018, Esmer:2021, Brown-Sevilla:2021}. These papers invoked one or more substellar companions as the primary cause of the observed ETVs. However, as new data became available, these models inevitably needed to be modified. 

After ruling out mass transfer, magnetic braking, and apsidal motion, \cite{Cakirli:1999} concluded that a third body with mass 23\Mjup{}and an orbital period of 19 yrs is the most probable cause. \cite{Ibanoglu:2004} also proposed a brown dwarf companion, with a minimum mass of 23\Mjup{} and orbital period of 18.8 yrs. \cite{Lee:2009} proposed two companions  with periods 15.8 and 9.1 yrs and masses of 19.2\Mjup{}and 8.5\Mjup{}respectively. 
\cite{Beuermann:2012b} added more ETV data, and showed that they deviated significantly from the solution proposed by \cite{Lee:2009}. They also showed that the solution proposed by Lee \etal\ is dynamically unstable. They proposed a new solution consisting of two brown dwarfs (14\Mjup{}+ 65\Mjup{}) that was stable over more than 10 Myr. However, \cite{Baran:2018} showed that the solution of Beuermann \etal\ does not fit more recent O-C data.

\cite{Horner:2012} independently made a stability analysis of the Lee et al. solution, and also found it highly unstable. Indeed, they searched a significant volume of substellar component parameter space, and could not find any stable solutions, concluding  "If the [\HWVir] system does host exoplanets, they must move on orbits differing greatly from those previously proposed". 
This failure to find a stable solution is reinforced by the  recent studies of \cite{Esmer:2021} and \cite{Brown-Sevilla:2021} using all previously published O-C observations along with new times of minima up to 2019.4. They tried a number of multi-planet models but  also failed to find a stable solution.

\subsection{HS0705+6700}

\cite{Drechsel:2001} first confirmed HS0705+6700 (V470 Cam, V= 14.7, P= 2.3 hr) as a post common envelope sdB binary. They fit their eclipsing timing observations with a linear ephemeris. 
\cite{Qian:2009} reported a cyclic change in orbital period and proposed  a substellar companion with orbital period of 7.15 years and mass of 40 \Mjup{}is responsible for the observed ETV. \cite{Beuermann:2012} observed an additional 28 epochs, with a similar solution:  A single brown dwarf with mass 31\Mjup{}and orbital period of 8.4 years.

In 2013, \cite{Qian:2013} fit a single brown dwarf model with a period 8.87 years and mass 32\Mjup, along with a positive quadratic term  $dP/dt = +9.8 \times 10^{-9} \textrm{\ days/year}$. They ruled out mass transfer and angular momentum loss due to gravitational radiation and magnetic braking as explanations of this period increase. \cite{Pulley:2015} showed that the more recent O-C timing data began to diverge from the Qian model. Later in 2018, they confirmed this departure by adding 65 more times of minima \citep{Pulley:2018}. 

\cite{Bogensberger:2017} proposed a new one planet model with a longer orbital  11.8 year period, eccentricity 0.38, and no quadratic term.  Recently, \cite{Sale:2020} added 20 additional ETV observations through mid-2018. They found a two component solution that fit all published O-C observations, but unfortunately it was highly unstable over a short timescale (1,000 yr) and hence not physically plausible.

\subsection{HS2231+2441}

HS2231+2441 (V = 14.2, P = 2.7 hr) is an eclipsing white dwarf + brown dwarf binary \citep{Almeida:2017} first reported by \cite{Ostensen:2007}.  They fit two years of  primary eclipse timing
observations with a linear (constant period)
ephemeris. \cite{Qian:2010}, using more extensive timing data, reported that the orbital period shows both a cyclic change and a period derivative.  They proposed a 14\Mjup{}component to explain the cyclic variation and magnetic braking to explain the period decrease. However, as  more timing observations became available, \cite{Lohr:2014} found that a linear function might provide a better fit to the O-C curve. \cite{Almeida:2017} reported eight more eclipse times and updated the linear ephemeris. Most recently, \cite{Pulley:2018} extended the timing database with 26 more timing minima. They compared the quadratic ephemeris by \cite{Qian:2010} and linear ephemeris proposed by \cite{Lohr:2014} concluded there is no statistically significant difference between the two.

\begin{table*}
	\centering
	\caption{Observing Summary}
	\label{tab:ObsSummary}
	\tabcolsep=0.2cm
    \renewcommand{\arraystretch}{1.3} 
	\begin{tabular}{lccccc} 
		\hline
		\hline
		Target       & V      & $N_{obs}$  & t(sec) &  Filter      &    Epochs\\
		\hline
		HS2231+2441  & 14.2   & 26         &  40    & Luminance  & 2018.9 -- 2020.8 \\
		HS0705+6700  & 14.7   & 40         &  40    & Luminance  & 2019.1 -- 2022.2 \\
		\HWVir       & 10.7   & 24         &  15    & Sloan r$'$ & 2018.9 -- 2022.1 \\
		\hline
	\end{tabular}
\end{table*}

\begin{table*}
    \caption{Mid-Eclipse Times  (BJD)}
    \label{tab:Minimum_times}
    \centering
    \begin{threeparttable}[b] 
    \tabcolsep=0.2cm
    \renewcommand{\arraystretch}{1.3} 
    \begin{tabular}{l |l l| c}
    \hline
    \hline
    \multicolumn{1}{c}{HS2231+2441} & \multicolumn{2}{c}{HS0705+6700} & \multicolumn{1}{c}{HW Vir}\\
    \hline
    2458471.69752(10) &    2458500.71590(3)  & 2459143.84436(5)  & 2458472.00981(2) \\
    2458473.57744(9)  &    2458504.82867(24) & 2459144.89662(6)  & 2458473.99404(4) \\
    2458617.89415(8)  &    2458505.59376(7)  & 2459145.85304(4)  & 2458607.75446(2) \\
    2458650.84983(9)  &    2458512.67178(8)  & 2459146.80949(6)  & 2458612.77339(2) \\
    2458991.90258(10) &    2458565.66003(3)  & 2459147.86162(4)  & 2458621.76088(2) \\
    2458992.89767(7)  &    2458589.66736(5)  & 2459149.87025(6)  & 2458622.69461(3) \\
    2458998.86963(8)  &    2458619.70026(7)  & 2459151.87879(7)  & 2458641.019582(41)* \\
    2459019.88131(6)  &    2458847.33965(5) * & 2459152.83526(7)  & 2458657.71049(3) \\
    2459020.87669(7)  &    2458985.74066(17) & 2459153.79168(6)  & 2458981.72376(3) \\
    2459021.87188(6)  &    2458987.74896(6)  & 2459156.85240(5)  & 2458981.84055(4) \\
    2459022.86704(9)  &    2458994.73099(8)  & 2459157.80886(8)  & 2458990.71110(2) \\
    2459023.86254(8)  &    2459122.89784(5)  & 2459158.86111(5)  & 2458992.69525(3) \\
    2459024.85779(8)  &    2459126.91491(6)  & 2459266.65475(11) & 2458996.78043(2) \\
    2459026.84836(6)  &    2459127.87140(6)  & 2459619.78262(3)  & 2458997.71417(2) \\
    2459138.65276(7)  &    2459128.92344(7)  &                   & 2458998.76469(2) \\
    2459139.64791(7)  &    2459129.88002(6)  &                   & 2459003.66701(2) \\
    2459140.64330(7)  &    2459131.88864(5)  &                   & 2459004.71749(2) \\
    2459141.63860(4)  &    2459132.84512(5)  &                   & 2459020.70807(3) \\
    2459142.63383(5)  &    2459133.89723(4)  &                   & 2459022.69231(4) \\
    2459143.62915(4)  &    2459134.85363(6)  &                   & 2459024.67650(3) \\
    2459144.62452(6)  &    2459136.86217(5)  &                   & 2459263.83482(3) \\
    2459145.61985(7)  &    2459138.87083(7)  &                   & 2459265.81903(3) \\
    2459146.61505(9)  &    2459139.82732(8)  &                   & 2459269.78746(3) \\
    2459147.61035(7)  &    2459140.87940(6)  &                   & 2459583.99637(1) \\
    2459151.59148(8)  &    2459141.83589(5)  &                   & \\
    2459152.58676(6)  &    2459142.88799(6)  &                   & \\
    \hline
    \hline
    \end{tabular}
    
    \begin{tablenotes}[flushleft]
    \small
    \item NOTE - $*$ Times of minima with asterisk are unpublished observations from David Pulley, John Mallet and the Altair Group. $a.$ HS0705+6700 - Dec 29th, 2019. $b.$ HW Vir - June 6th, 2019.
    \end{tablenotes}
    \end{threeparttable}
\end{table*}

\begin{table*}
    \caption{Binary star physical properties }
    \centering
    \begin{threeparttable}[b] 
    \label{tab:Binary_props}
    \tabcolsep=0.2cm
    \renewcommand{\arraystretch}{1.3} 
    \begin{tabular}{l cccccccc}
        \hline
        \hline
        Binary       & Type     & M$_{pri}$  & M$_{sec}$ &  $R_{pri}$      &   $R_{sec}$ & $L_{sec}$ &  $P_{orb}$ [d] &  Refs  \\
        \hline
        HS2231+2441 & WD+BD & 0.288 & 0.046 &  0.165 & 0.086 & 0.0009 & 0.1106  & \cite{Almeida:2017} \\
        HS0705+6700 & sdB+M4-5 & 0.483 & 0.134 & 0.230 & 0.186 & 0.0022 & 0.0956 & \cite{Drechsel:2001} \\
        HW~Virginis & sdB+M4-5 & 0.485 & 0.142 & 0.183 & 0.175 & 0.0025 & 0.1167 & \cite{Lee:2009} \\
        \hline
    \end{tabular}

    \begin{tablenotes}[flushleft]
      \small
      \item NOTE - $a.$ All masses, radii, and luminosities are solar units. $b$. \cite{Almeida:2017} presents two mass solutions for HS2231: We have listed solution 2. $c$. \cite{Baran:2018} derived a primary mass 0.26 M$_\odot$ for HW~Vir, but argue that it is probably an underestimate.
    \end{tablenotes}
    \end{threeparttable}
\end{table*}

\section{Observations}

\begin{figure}
\centering
\includegraphics[keepaspectratio=true,width=\linewidth]{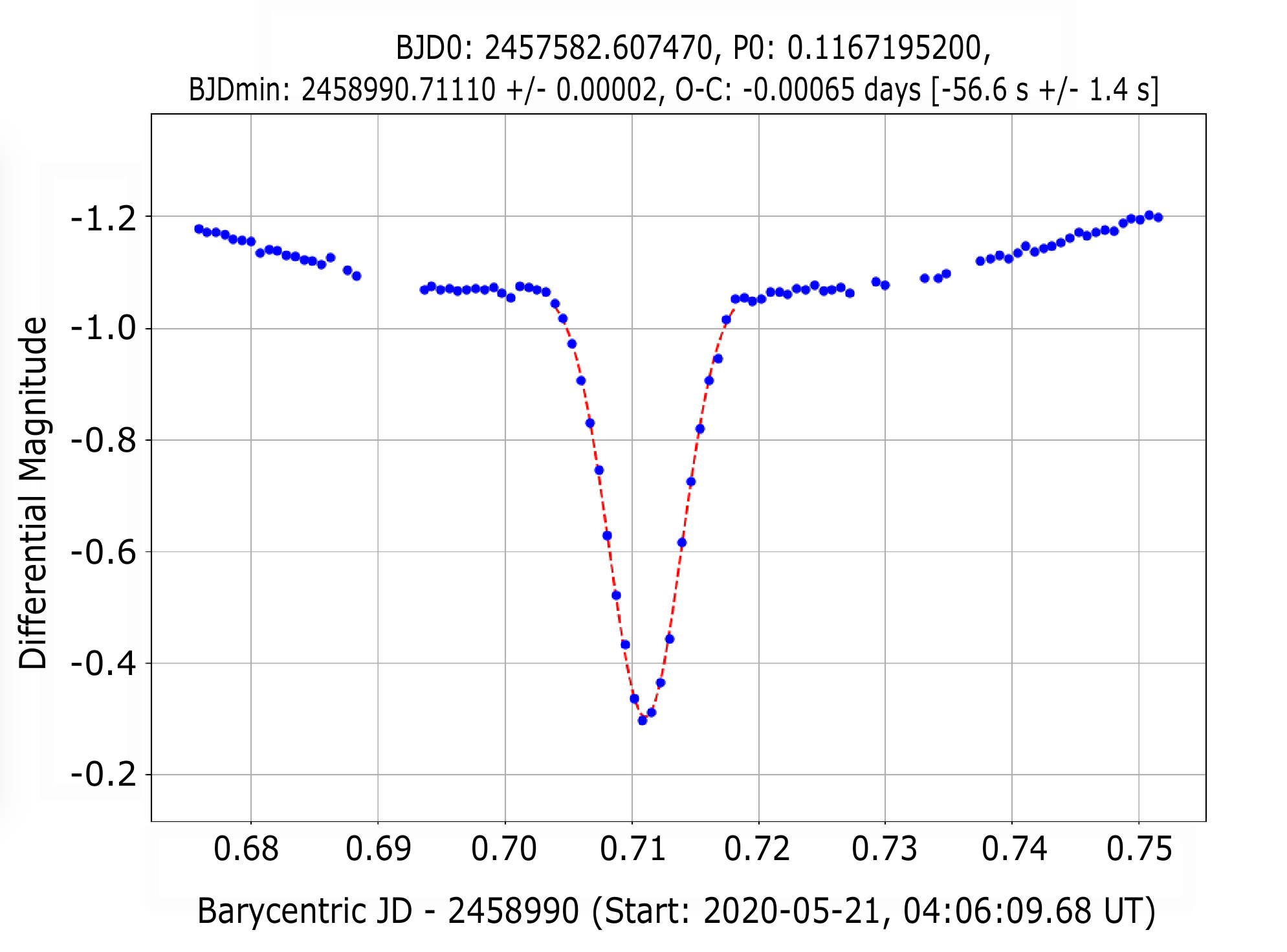}
\caption{Sample HW\ Vir light curve taken at the Iowa Robotic Observatory on 21 May 2020 from 04:06 --- 06:01 BJD, with fitted Gaussian profile to determine time of minima [red dashed line].}\label{fig:sample-lc}
\end{figure}

Our study is an independent analysis of previously published eclipse times and new timing data obtained with the 0.5m telescope at the Iowa Robotic observatory\footnote{\url{http://astro.physics.uiowa.edu/iro}}, located at Winer Observatory\footnote{\url{https://winer.org/}} in southeastern Arizona. We used a SBIG 6303e CCD camera (2048x3072 x 12 micron pixels) with an field of view 28x18 arcmin. We report 90 new times of minimum for three PCEB systems at primary eclipse between December 2018 and February 2022 (Table \ref{tab:ObsSummary}). A luminance filter was used for HS2231+2441 and HS0705+6700, and an exposure time of 40 sec was adopted. A Sloan r$'$ filter with an exposure time of 15 sec was used for the brighter object HW Vir.

After the normal CCD image calibration including dark frame subtraction and flat fielding, we created light curves using airmass-corrected differential photometry. Both HS 2231+2441 and HS0705+6700 are typical of HW Vir-like eclipsing systems, their light curves show a strong reflection effect with a very deep primary and a shallow secondary minimum. We fit a Gaussian profile to the primary eclipse light curve using a nonlinear least-squares (Levenberg-Marquardt) algorithm and determine the observed mid-eclipse times. A sample light curve for HW~Vir with associated Gaussian profile fit is shown in Fig.~\ref{fig:sample-lc}.
We evaluated the uncertainty in time of eclipse minimum by a Monte Carlo method consisting of a series synthetic eclipse light curves with randomly-chosen times of minima and different exposure times. This procedure is described in detail in Appendix 1. Table \ref{tab:Minimum_times} lists the eclipse timing observations and uncertainties for all three PCEBs.

\section{Data Analysis}

\subsection{Data selection}

Our eclipse timing dataset compromised 90 new time of primary eclipse minima (Table~\ref{tab:Minimum_times}) together with previously published times from the literature. Before use, we made a few modifications.
\begin{enumerate}

\item For HW Vir and HS2231+2441, we excluded a portion of archival SuperWASP times of minima that were rejected by \cite{Lohr:2014}'s  outlier methodology and by the size of their uncertainties. In particular, we did not include \cite{Lohr:2014}'s listed extra times of minima due to their precision, and discarded timing data for which uncertainties are between 8 sec $\sim$40 sec.  We also omitted the individual superWASP times of minima for HS2231+2441, since they had large ($\sigma > 20s $) uncertainties. 
\item We also removed O-C eclipse timings obtained from the Mercator observation in \cite{Baran:2018} for HW Vir since they deviate from the overall O-C trend line by 50 seconds. We also excluded outliers that had O-C values deviating by more than 400 seconds from the overall O-C trend.

\end{enumerate}

\subsection{Light Travel Time Model}
For a binary system with orbiting substellar components, the predicted primary mid-eclipse time as a function of cycle number $E$ can be written as 

\begin{equation}
T(E) = T_0 + P_0E + \alpha E^2 + \sum_i \tau_i(E)
\end{equation}

where $T_{0}$ is the reference epoch, $P_{0}$ is the linear eclipse period, where $\tau_i$ are the time-dependent contributions from each perturbing body. and
\begin{equation}
\alpha = \frac{1}{2}\frac{dP}{dt}P_0,
\end{equation}
is a quadratic term can would result from non-planetary effects, such as apsidal motion, mass exchange, magnetic braking, or gravitational radiation.
We used the usual formulation of  \cite{Irwin:1952, Irwin:1959} to calculate the
time signature of perturbing bodies on eclipse timing as a function of each body's orbital properties.

\subsection{Orbit stability}
A physically credible model must not only fit all the current and previously published ETV datasets, but it is also subject to the prior constraint that the substellar components must be stable to orbital instabilities over timescales at least comparable to the lifetime of the PCEB phase ($\sim$ 100 Myr) \citep{Han:2002,Han:2003,Zorotovic:2011}. We used the Hill stability criterion \citep{Gladman:1993, Chambers:1996} to predict whether any two components in orbit about a massive central body will avoid mutual close encounters over many orbits. 

For circular orbits, stability occurs if the semi-major axis difference between pairs of components exceeds a critical distance $\Delta$ given by \citep{Chambers:1996},

\begin{equation}
    \Delta \simeq 2\sqrt{3}\cdot\  R_H
\label{eqn-hill}
\end{equation}

where $R_H$ is the  Hill radius,

\begin{equation}
   R_H = \frac{a_1+a_2}{2}  \left[\frac{(m_1+m_2)}{3M} \right]  ^ \frac{1}{3}
\end{equation}

where $m_1,m_2$ and $a_1, a_2$ are the masses and semi-major axes of the components respectively, and $M$ is the mass of the central body.

For non-circular orbits, a more general stability criterion is \citep{Hasegawa:1990}

\begin{equation}
\Delta^2 = \left[12 + \frac{4}{3}K^2e^2\right]\ R_H^2
\end{equation}

where
\begin{equation}
K = \left[\frac{(m_1+m_2)}{3M} \right]  ^ {\frac{1}{3}},
\end{equation}
and $e$ is the larger eccentricity. For low eccentricity orbits and $K\ll 1$, we can expand the second term to obtain

\begin{equation}
\Delta = 2\sqrt 3\ R_H\ \cdot\left(
1+\frac{K^2e^2}{18}
\right).
\end{equation}

The second term in parentheses is much less than 1 for the components considered here ($K\ll1$) so the simpler expression equation~\ref{eqn-hill} was used in our calculations. 

We can define a dimensionless parameter $\beta$ that characterizes each planet pair spacing in terms of the normalized mutual Hill radii as \citep[cf. ][]{Smullen:2016}

\begin{equation}
\beta = \frac{\left|a_2-a_1\right|}{\Delta}
\label{eqn:hill_normalized}
\end{equation}

Stable component pairs correspond to $\beta>1$ while unstable pairs have $\beta<1$. 

For systems with more than two substellar components, instability can occur even if all components lie outside the Hill radius of their nearest companion, although the timescale for close encounters increases as the initial component separations become larger \citep{Chambers:1996}. Hence, we used the Hill parameter as an optimization  prior but ran numerical orbit simulations to test orbit stability after optimization, as described below.

\subsection{Model Fitting Scheme}

\begin{figure}
\centering
\includegraphics[keepaspectratio=true,scale=0.45]{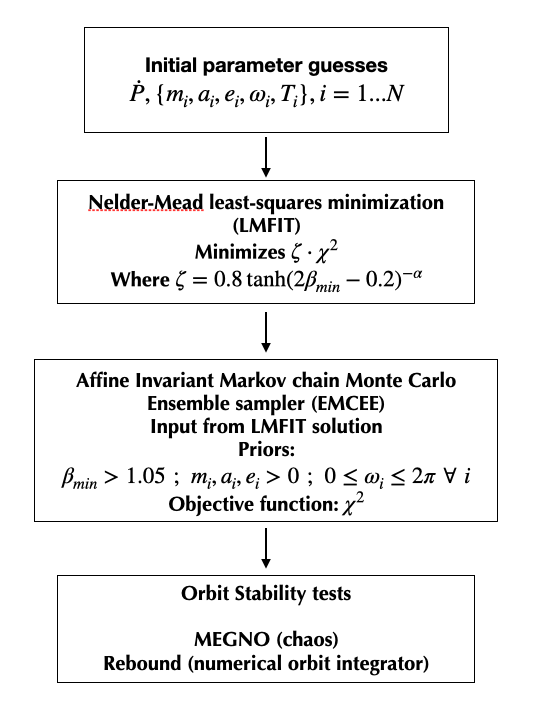}
\caption{Flow chart of data analysis.}
\label{fig:flowchart}
\end{figure}

The model fitting scheme comprised four steps (Fig.~\ref{fig:flowchart}):  
\begin{enumerate}
    \item Create an initial model of between one and four sub-stellar components, with multi-component models spaced at least one Hill radius apart (i.e., $\beta>1$ for all pairs),
    \item Search for the minimum of an objective function proportional to  the product of the summed $\chi^2$ of the model vs. O-C time series and an ansatz 'penalty' function that sharply increases when the smallest Hill parameter $\beta$ becomes less than 1.0 for any component pair,
    \item Use Bayesian inference to solve for the final fitted parameters using a Markov chain Monte Carlo (MCMC) sampling, a weighted least-squares objective function, and an orbit stability prior ($\beta>1$), 
    \item Run orbit stability tests using both the N-body orbital integration package REBOUND \citep{Rein:2012,Rein:2015} and the chaotic indicator program MEGNO\ \citep{Cincotta:2000, Hinse:2010}. 
\end{enumerate}

We first chose the number of substellar companions to model. Examination of the O-C plot for both HS0705+6700 and HW Virginis indicates that a large number of components, although certainly possible, is not justified by the data, since the model parameters are already highly degenerate for few component models, as we will see. Hence we restrict our model selection to one through four component models. Each component is characterized by five parameters: mass, semi-major axis, eccentricity, longitude of periapsis $\tilde{\omega}$, and time of periastron passage . We restricted the model to co-planar orbits for the sake of computational simplicity.
We also added a period derivative parameter to account for the possible contribution of non-orbiting components, such as apsidal motion or gravitational radiation.

\begin{figure}
\centering
\includegraphics[keepaspectratio=true,width=\linewidth]{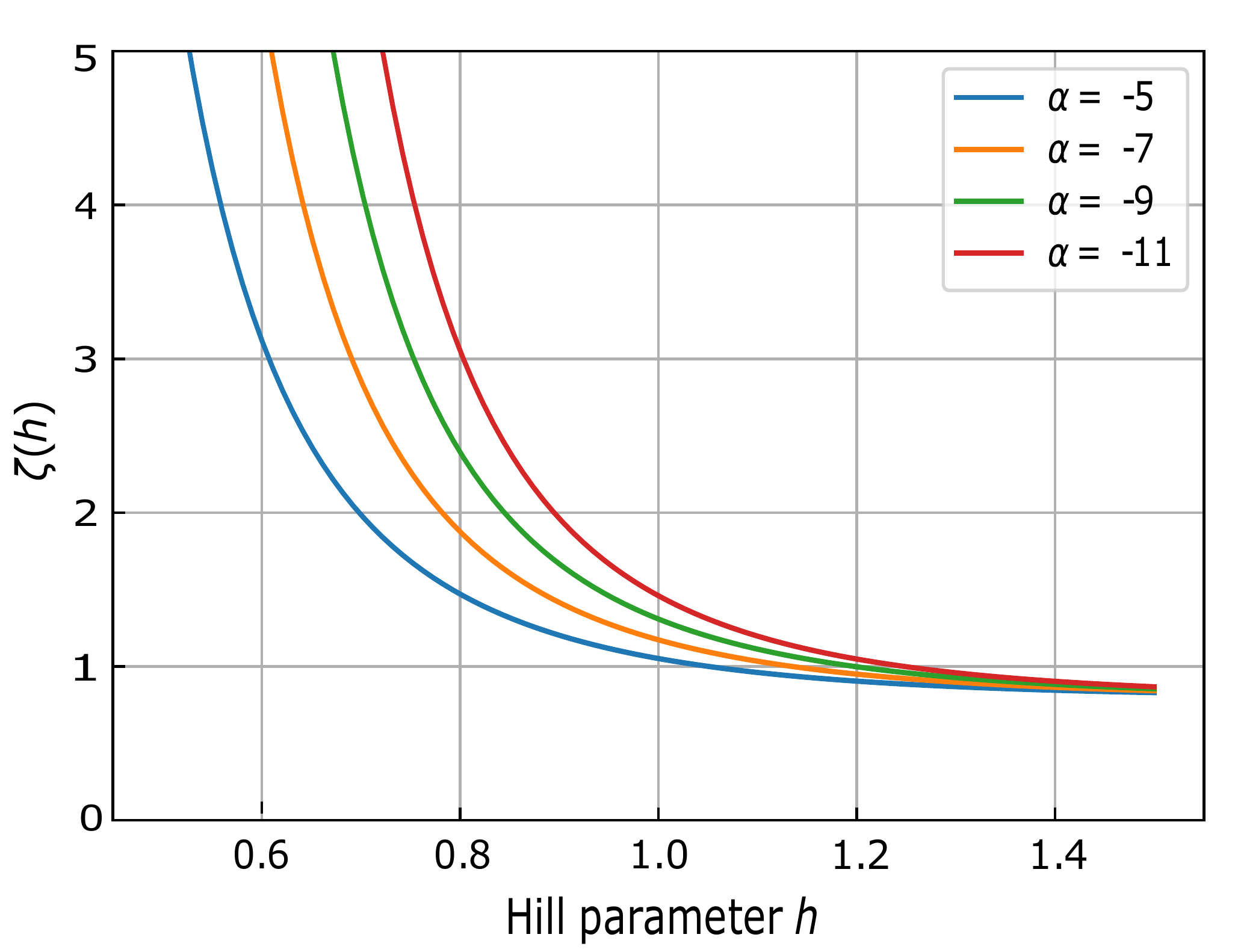}
\caption{Hill penalty function used in prior function (see text). (Eqn.~\ref{eqn:ansatz}).}
\label{fig:hill_wt}
\end{figure}

We started model fitting using a downhill simplex (Nelder-Mead) minimization   (Python library LMFIT\footnote{\url{https://lmfit.github.io/lmfit-py}} to fit the O-C curve. However, instead of minimizing the sum-squared differences between the O-C data and the model, we incorporate the orbit stability constraint by minimizing the product,

\begin{equation}
\zeta\cdot\chi^2
\end{equation}
Where we define an empirical Hill stability function
\begin{equation}
\zeta = 0.8\ \rm{tanh}\left(2\beta_{min}-0.2\right)^{-\alpha}
\label{eqn:ansatz}
\end{equation}
where the Hill parameter $\beta_{min}$ is the smallest normalized Hill parameter (equation~\ref{eqn:hill_normalized}) of all component pairs. This stability function asymptotes to unity for $\beta>1$ but increases sharply for $\beta<1$  (see Fig.\ref{fig:hill_wt}).

The resulting model was used as a starting point for Bayesian inference modeling. We used the EMCEE \citep{Foreman:2013} Python implementation\footnote{\url{https://github.com/dfm/emcee}} with 4096 walkers and 4000 iterations. The likelihood function was summed-squared difference between the model and O-C data weighted by the data uncertainties. We used three priors: (i) stability parameter $\beta>1.05$, (ii) all masses and semi-major axes positive-definite, and (iii) eccentricities restricted to the range $0<e<1$.  

Using a model fitting scheme with a Bayesian methodology as the final step has two important advantages over the more commonly-used frequentist least-squares minimization approach. First, the orbit stability constraint (Hill parameter) can be incorporated naturally as a Bayesian prior without the {\it ad hoc} ansatz function we used in the LMFIT minimization. Second, the Bayesian posterior distribution provides a more comprehensive description of the parameter uncertainties and correlations. 

For each model, we tested dynamical stability by numerically integrating the component orbits using WHFAST, a Wisdom-Holman symplectic integrator \citep{Rein:2015a}. We used a time step of 36.5 days and integrated for 10 Myr for each model. In addition, we tested for chaotic behavior using the MEGNO algorithm (Mean Exponential Growth factor of Nearby Orbits, \cite{Cincotta:2000}) also with a time interval of 10 Myr. Both WHFAST and MEGNO were called from the Python package REBOUND \citep{Rein:2012} which also provided convenient tools for plotting orbits and time histories of orbital elements.

\section{Results}

\subsection{HS2231+2441}
\begin{figure}
\centering
\includegraphics[keepaspectratio=true,width=\linewidth]{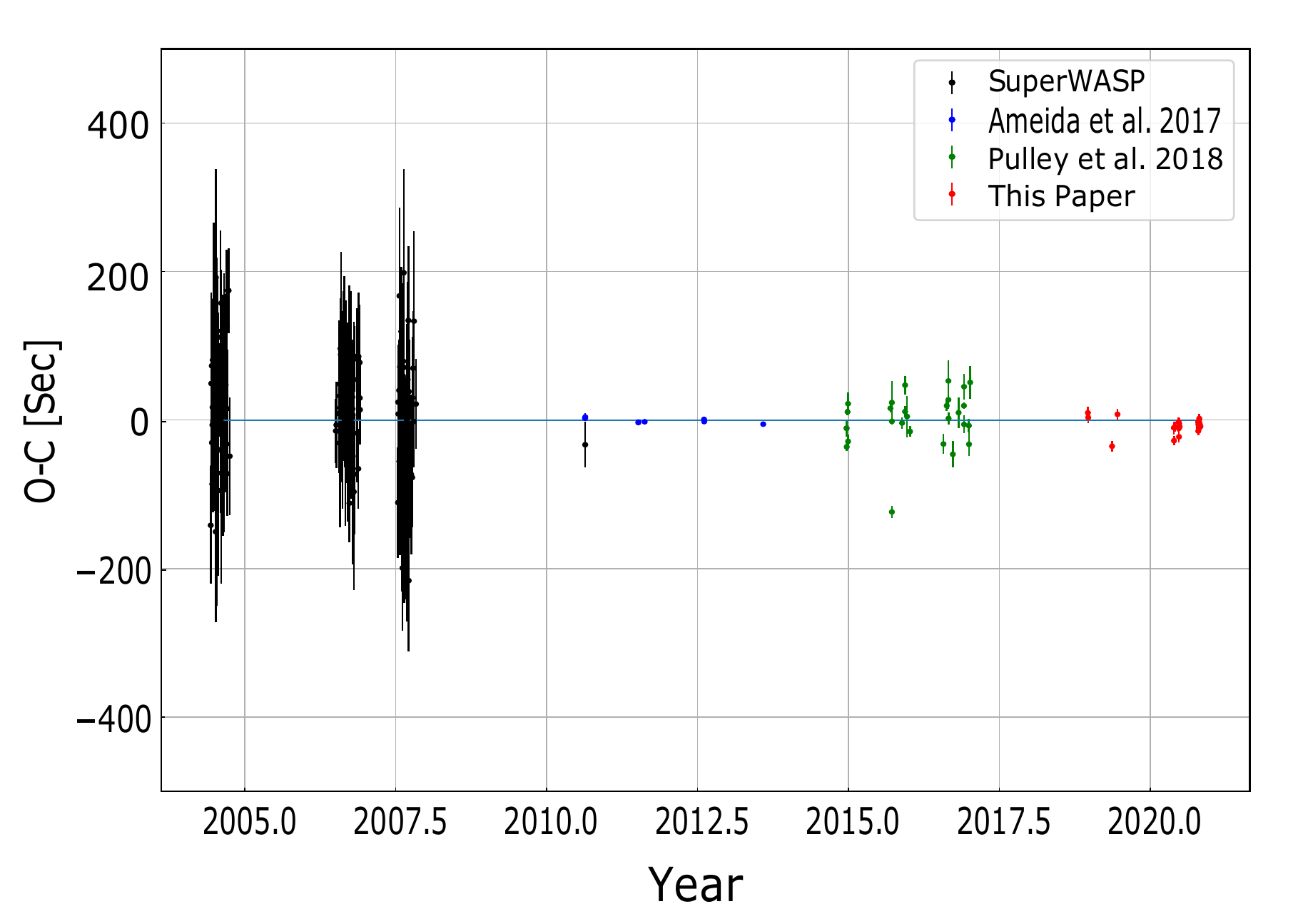}

\caption{HS2231+2441 O-C times of primary minima computed using the linear ephemeris given in Equation \ref{eqn-hs2231-ephemeris}. Observations shown as blue data points are times of minima from \protect\cite{Almeida:2017}, while green points are from \protect\cite{Pulley:2018}, and red points are new times of minima(Table~\ref{tab:Minimum_times}).}
\label{fig:hs2231-oc}
\end{figure}

After adding 26 additional primary eclipse timing observations (listed in Table~\ref{tab:Minimum_times}) to previously published observations, we fit the linear ephemeris of \cite{Almeida:2017} with a slightly changed orbital period,

\begin{equation}
{\rm BJD}_{min} = 2455428.76185(10) + 0.11058786(10)\cdot E
\label{eqn-hs2231-ephemeris}
\end{equation}
 The O-C diagram with all
of the available data is shown in Figure \ref{fig:hs2231-oc}.

We find that there is no statistically significant period variation for HS22315+2441 spanning 16 years, from 2004 to 2020, but we cannot rule out the variations smaller than 15 seconds. A linear fit without a orbiting planet is consistent with the full data set.

\subsection{HS0705+6700}

\begin{figure*}
\centering
\includegraphics[scale=0.8]{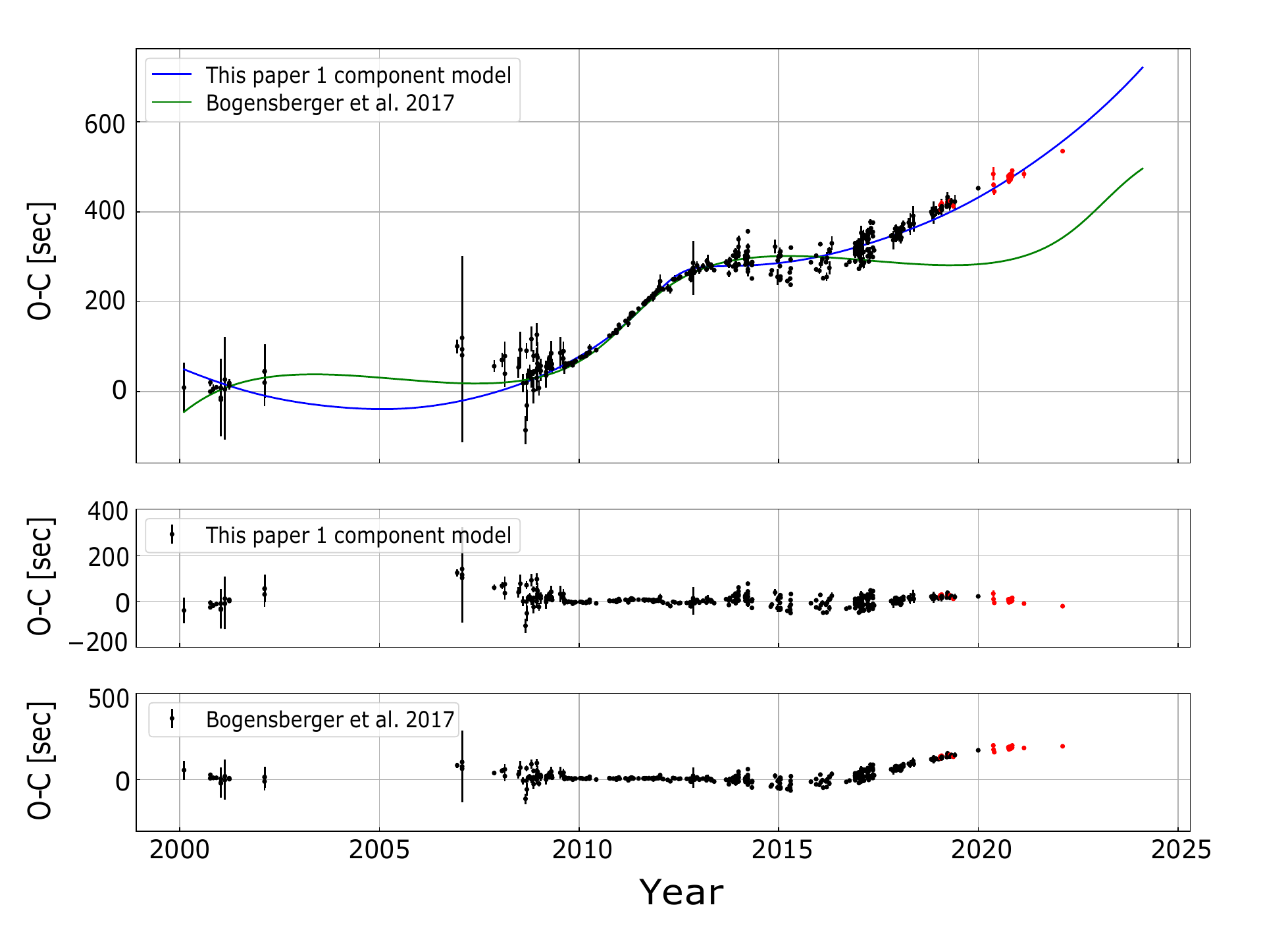}
\caption{{\it Upper panel}: HS0705+6700 O-C plot with one component model (blue line) and one component model (green line) of \citealt{Bogensberger:2017} overlaid. The black data points are from the literature, while the red data points are from Table~\ref{tab:Minimum_times}. {\it Lower panels:} O-C differences between observed O-C and model predictions.}
\label{fig:hs0705_oc}
\end{figure*}

\begin{figure*}
\centering
\includegraphics[scale=0.75]{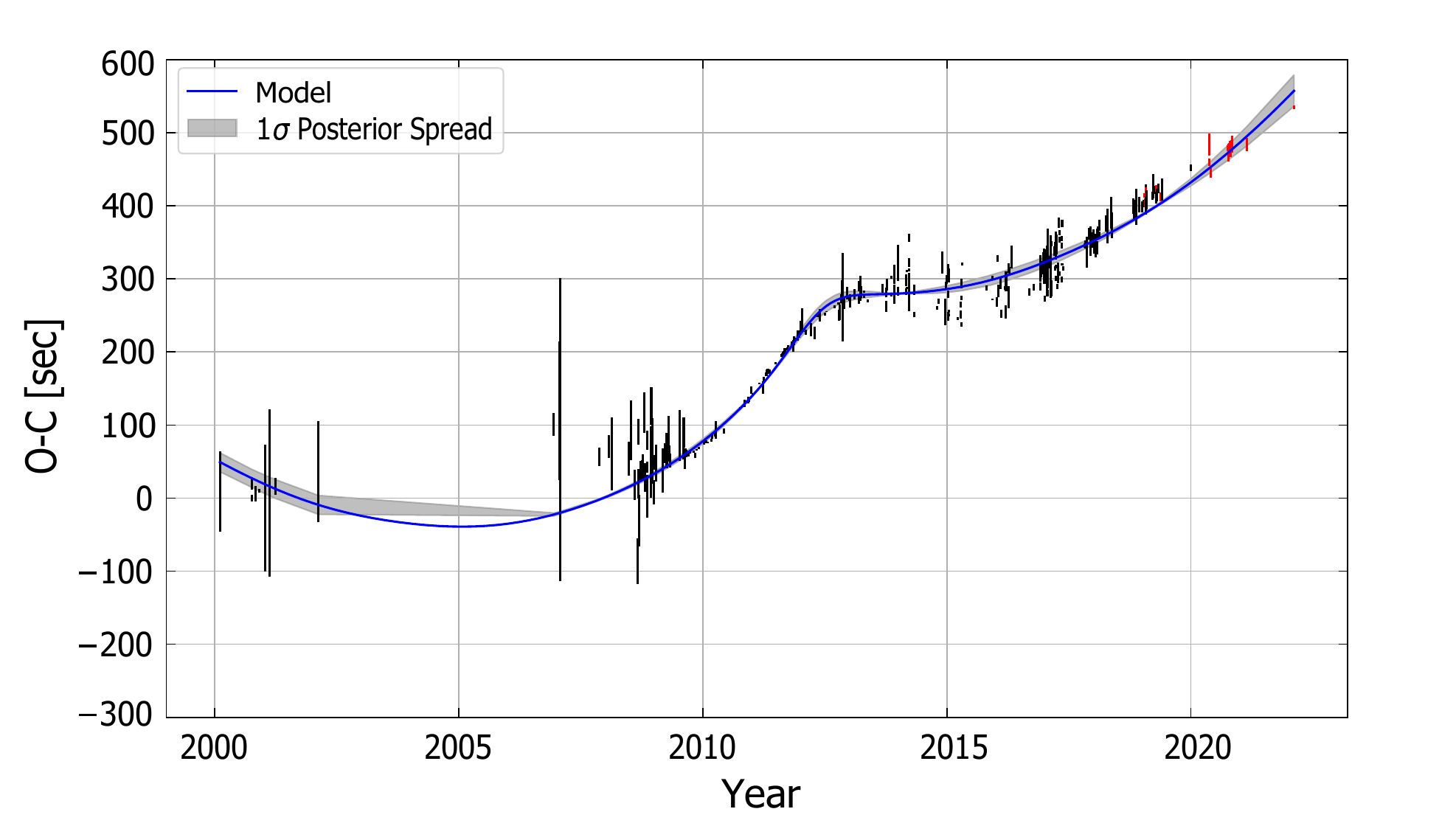}
\caption{HS0705+6700 O-C plot of one component model. The gray shaded area is the $\pm1\sigma$ spread of the EMCEE posterior samples distribution. The black data points are from the literature, while the red data points are from Table~\ref{tab:Minimum_times}.}
\label{fig:hs0705_sigma}
\end{figure*}

\begin{figure*}
\centering
\includegraphics[scale=0.85, angle=0]{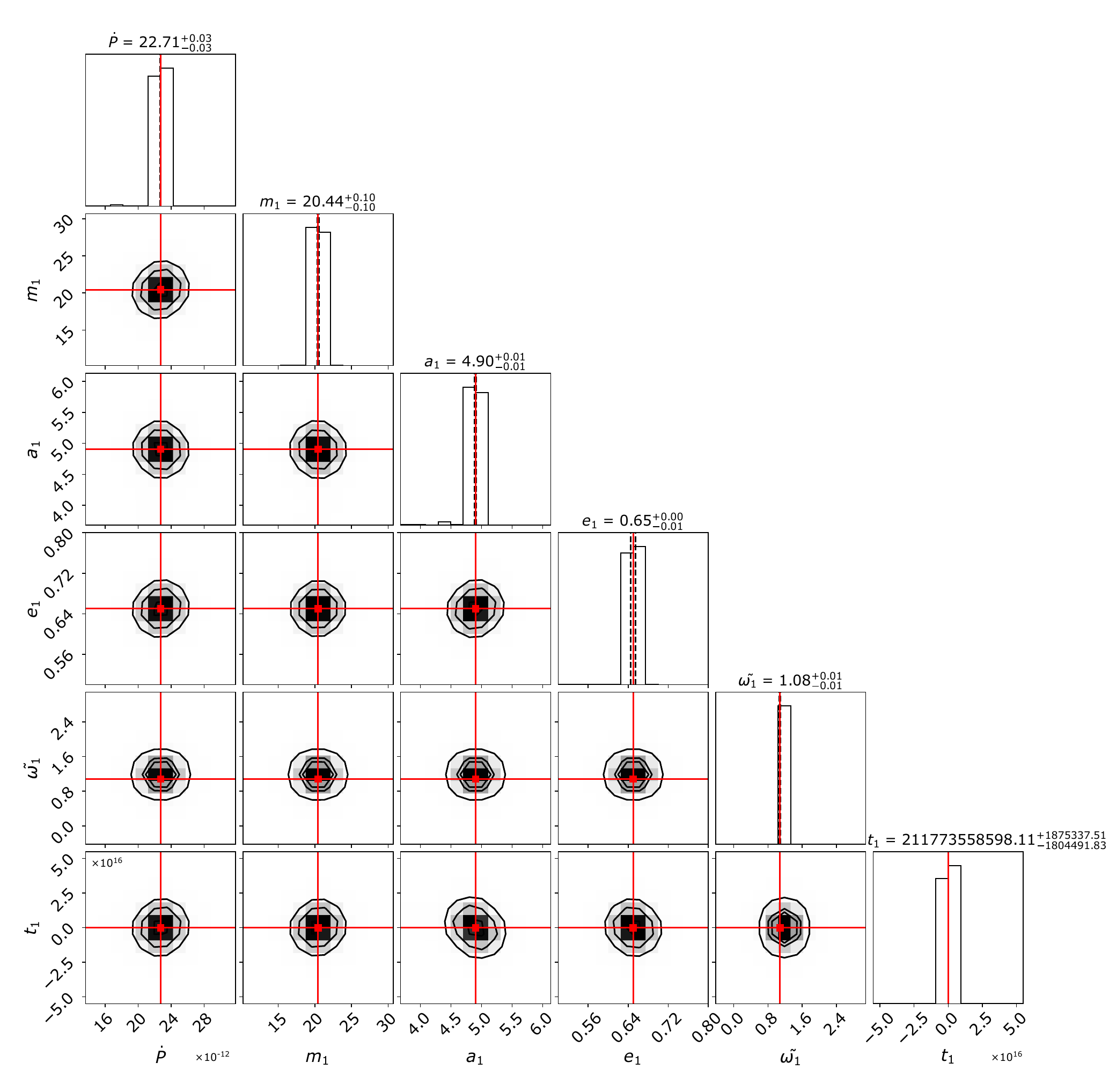}
\caption{Joint posterior model parameter probability distributions for 1-component model for HS0705+6700, as listed in Table~\ref{tab:hs0705_1comp_params} respectively. Our model parameters converge to the median values, shown in red, and contours 16\%, 50\%, and 84\% quantiles. Note the time of periastron passage is in shown in seconds. }
\label{fig:hs0705_corner}
\end{figure*}

Figure \ref{fig:hs0705_oc} Shows the O-C timing residuals versus epoch computed using linear ephemeris,

\begin{equation}
{\rm BJD_{min}} = 2451822.760509 + 0.095646609  \cdot E
\end{equation}

along with our one component model (solid blue line) and the one-component model of \citealt{Bogensberger:2017} (solid green line). Table~\ref{tab:hs0705_1comp_params} lists the model parameters of our model. 

We also found a best-fit two component model with masses 21.8\Mjup{} and 1\Mjup{} and semi-major axes 4.9 AU and 37.3 AU, respectively. The two component model was stable for at least $10^{7}$ years as verified by REBOUND and MEGNO simulations, but the fit to the timing data was no better than the one component model. Therefore, we do not discuss this model further.

It is clear that the previously published model of Bogensberger et al. deviates strongly from the observed timing data starting shortly after the publication of the model. As previously noted, this discouraging trend has been seen for several other eclipsing sdB binary exoplanet detection. Also, we did not plot the recent two component solution proposed by \cite{Sale:2020} since we could not determine their ephemeris or interpret orbital parameters of their model from their paper. 

An expanded view of the O-C fits for our model is shown in Fig.~\ref{fig:hs0705_sigma}, along with the $\pm1\sigma$ posterior spread from the EMCEE solutions. Fig.~\ref{fig:hs0705_corner} shows the two-dimensional projections of the sample covariances for our one component model. They demonstrate that our model have well-defined parameters with small fractional uncertainties. 

\begin{table*}
    \caption{HS0705+6700 One Component Model}
    \label{tab:hs0705_1comp_params}
    \centering
    \tabcolsep=0.2cm
    \renewcommand{\arraystretch}{1.5} 
    \begin{tabular}{l r r r}
        \hline
        \hline
        \multicolumn{1}{c}{Parameter} & \multicolumn{2}{c}{Fitted Values} & \multicolumn{1}{r}{Unit}\\
        \hline
        \multicolumn{4}{c}{Inner Binary} \\
        
        \hline 
        \vspace{2pt}
        T$_0$     & \multicolumn{1}{r}{2451822.760509}     &     & BJD       	 \\
        \vspace{2pt}
        P$_0$     & \multicolumn{1}{r}{0.095646609}       &      & day	         \\
        \vspace{2pt}
        $\dot{P}$  & \multicolumn{1}{r}{${22.71_{-0.03}^{+0.02}}$} &	& $10^{-12}$\ s/s  \\
        
        \hline
        \multicolumn{4}{c}{ Substellar Component} 	\\
        
        \hline
        \vspace{2pt}
        $P$ 	     & $13.63_{-0.06}^{+0.05}$      &  & yr    \\
        \vspace{2pt}
        $K$          & $74.86_{-0.20}^{+0.22}$ 	     & & sec	 \\
        \vspace{2pt}
        $asin(i)$	 & $4.91_{-0.01}^{+0.01}$       & & AU	 \\
        \vspace{2pt}
        $e$	         & $0.65_{-0.01}^{+0.01}$   & &      \\
        \vspace{2pt}
        $\tilde{\omega}$	 & $1.08_{-0.01}^{+0.01}$    &  & radian \\
        \vspace{2pt}
        Min. Mass	 & $20.45_{-0.10}^{+0.10}$     & & \Mjup  \\
        \vspace{2pt}
        $T$	 & $2451081.52413_{-17.52}^{+24.26}$      & & day  \\
        
        \hline
    \end{tabular}
\end{table*}

\subsection{HW Virginis}

\begin{figure*}
    \centering
    \includegraphics[angle=0,width=1.0\textwidth]{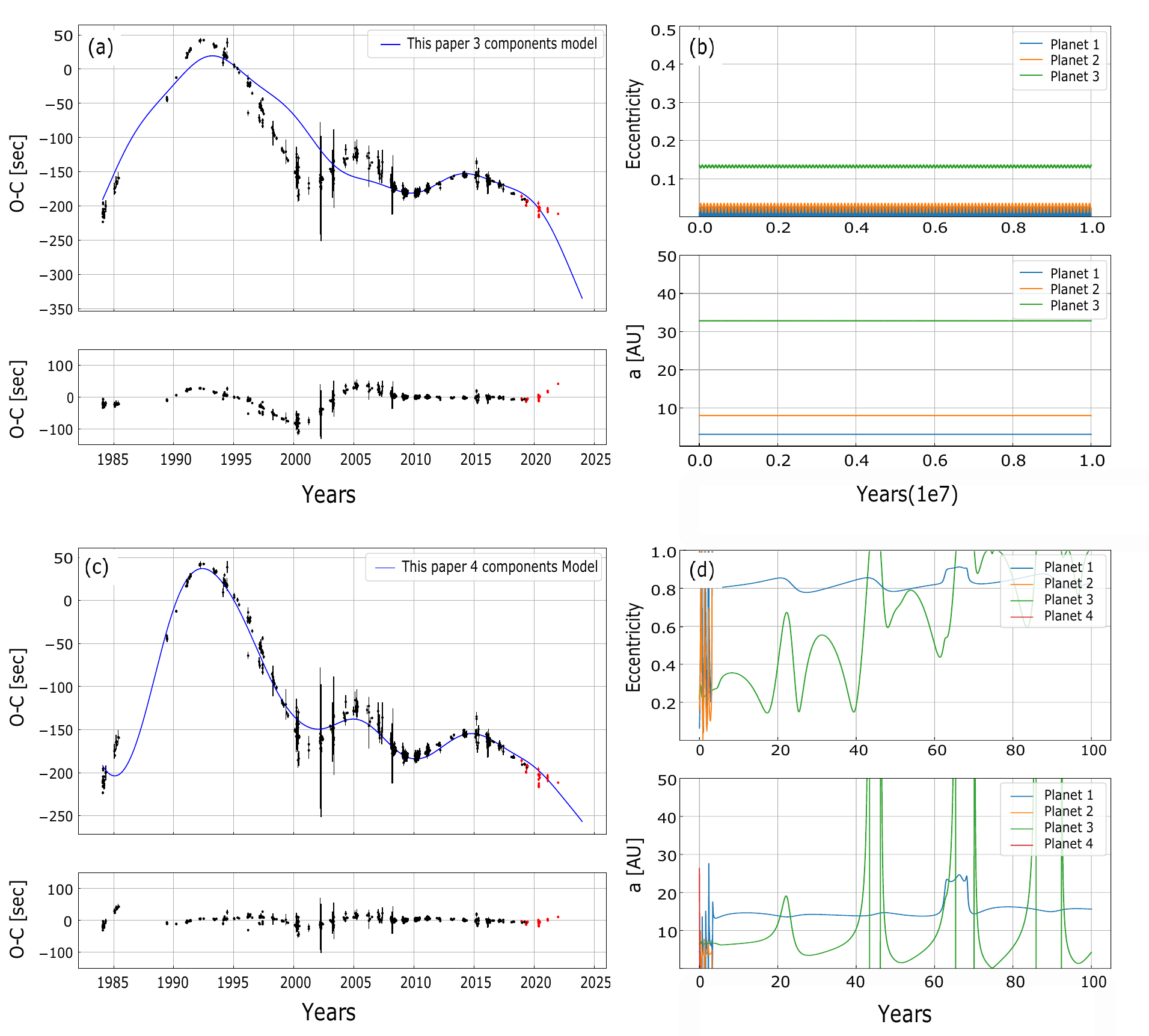}
    \caption{\HWVir O-C times of primary minima computed using the linear ephemeris given in Equation~\ref{eqn:hwvir-linear}. Black points are archival times of
minima from the literature, red points are  new times of minima from Table~\ref{tab:Minimum_times}, and blue lines are model predictions. 
(a) Best-fit 3-component model subject to the constraint that the model is stable i.e., all pairwise component normalized Hill parameters exceed 1.05. 
(b) Eccentricity and semi-major axes of all three components over 10 Myr. This solution is stable but is a very poor fit ($\chi^2_r\approx 93$) to the timing data.
(c) Best-fit 4-component model.
(d) Eccentricity and semi-major axes of all four components over 100 yr. 
The four component model provides a much improved fit ($\chi^2_r\approx 19$) but is highly unstable.
We conjecture that the observed ETV's may be result from a combination of gravitational perturbations by substellar companions and a Applegate-Lanza mechanism.}
\label{fig:oc_hwvir}
\end{figure*}

Figure \ref{fig:oc_hwvir} Shows the O-C timing residuals versus epoch using the linear ephemeris
\begin{equation}
{\rm BJD_{min}} = 2450280.28596 + 0.116719519 \cdot E
\label{eqn:hwvir-linear}
\end{equation}
On the right of each panel are plots of eccentricity and semi-major axes of all model components vs. epoch as computed by numerical integration using REBOUND with time steps of 0.1 yr. The plots also show two model fits: 
\begin{itemize}
\item The upper panel shows a three component model computed using the minimization scheme described in section 3, with a Bayesian prior  $\beta>1.05$ for all component pairs, where $\beta$ is the normalized Hill constraint. This solution is stable for at least 10 Myr, but is a very poor fit to the timing data ($\chi^2 \approx 93$).
\item The lower panel shows a  4-component model, also computed using the minimization scheme in section 3, but without the Hill stability prior. This provides a much better fit ($\chi^2\approx 19$) but is highly unstable over short time internals (one hundred years).
\end{itemize}
We do not list the orbital elements for either model since they are implausible: They are either very poor fits to the timing data (3-component stable model) or are highly unstable (four-component model). 
This unfortunate scenario (acceptable O-C model fits are invariably unstable) was also found by \citep{Brown-Sevilla:2021} who analyzed a series of best-fit six multi-component  models for HW Vir. In the next section we consider possible alternatives to pure substellar component models.

\section{Alternative explanations for eclipse timing variations}

As mentioned previously, eclipse timing variations can be caused by mechanisms other than gravitational perturbation by orbiting circumbinary masses. These include interbinary mass transfer, apsidal motion, and changes in the moment of inertia caused by magnetic activity on one of the components (Applegate mechanism). 

For sdB binaries, mass transfer is ruled out since they are detached systems. Apsidal precession can result from (general relativistic) gravitational radiation and by tidal effects caused by an asymmetric mass distribution in one (or both) components. \cite{Lee:2009} has shown that gravitational radiation for the sdB binary HW Vir, and by extension other similar systems including HS0705+6700, would result in time variations at about 2 dex smaller than observed. 

Ruling out apsidal effects is more problematic since they depend on the tidal distortion of the internal mass distribution (via the internal structure constant $k_2$, \citealt{Sirotkin:2009}) which is not well-characterized for these systems. \cite{Almeida:2020} recently considered tidal and rotation apsidal effects on eclipse timing variations for the PCEB GK~Vir, They found that apsidal is less likely than substellar component model, but could not rule it out as a contributing factor. In addition, apsidal effects will produce strictly periodic signatures in O-C plots, which is not observed for the binaries considered in this paper. Hence, we do not consider apsidal effects further, but examine the possible contribution of Applegate-type mechanisms.

Applegate \citep{Applegate:1987, Applegate:1992} proposed that orbital period changes in close binaries may result from a variable quadrupole moment produced by magnetic activity in the outer convection zone of one of the stars. For sdB binaries, this is likely to be the low-mass secondary, which is fully convective. Although we are not aware of any direct evidence for magnetic fields in any sdB secondary star, there is ample evidence for strong (kiloGauss) surface fields in many late-type (M0-M5) main sequence single stars similar to the secondaries in sdB binaries \citep[e.g.,][and references therein]{Kochukhov:2021}. Also, tidal locking will ensure that the rotation of the secondary is very rapid, enhancing the prospects for a robust magnetic dynamo.

The original Applegate mechanism assumed a time-dependent gravitational quadrupole moment modulated by the magnetic activity cycle in an outer thin shell of the active star. An increasing quadrupole moment results in a stronger gravitational field, driving a decreasing orbital radius and consequent reduction of the binary period. However, it was subsequently found that energy required to drive quadrupole moment variations of the secondary star that could account for the ETV’s was inconsistent with observed luminosity changes in many close binaries \citep{Volschow:2016}. 

Several recent papers have analyzed modifications to the original Applegate model using more complex 2-shell models, generalizing the magnetohydrodynamic effects of internal magnetic fields, and considering different types of dynamo models  
\citep[e.g.,][]{Lanza:1998,Lanza:2006, Navarrete:2018, Volschow:2018, Navarrete:2020}. Here we consider the recent model of \cite{Lanza:2020} which is based on angular momentum exchange between the spin of the active component and the orbital motion. Spin–orbit coupling is produced by a non-axisymmetric component of the gravitational quadrupole moment of the active star caused by an assumed persistent non-axisymmetric internal magnetic field. 

\subsection{Hybrid models with substellar components and Applegate-type modulation}

As we have seen, the best-fit stable orbit model fits for both HS0705 and HW~Vir are not fully in agreement with the observed O-C data, with residual average differences of order 10 sec and 20 sec over 10-20 year timescales for HS0705 and HW ~Vir respectively  (Figs.~\ref{fig:hs0705_oc},\ref{fig:oc_hwvir}).  We now consider the question of whether the ETV variations may result from a {\it combination} of substellar component perturbation and Applegate-Lanza effects. 

The change in fractional angular velocity $\Omega$ of the secondary is related to the observed fractional period changes by \citep[][eqn. 56]{Lanza:2020}
\begin{equation}
\frac{\Delta\Omega}{\Omega} = -\frac{ma^2}{3I_s}\frac{\Delta P}{P}
\end{equation}
Where $m$ in the reduced mass of the binary, $a$ is the semi-major axis of the binary, $I_s$ is the moment of inertia about the active (secondary) star spin axis, and $\Delta P/P = \Delta t/T$ is the fractional change in the period, or equivalently the O-C deviations $t$ over the timescale $T$ of the O-C data . The consequent change in rotational energy is
\begin{equation}
\Delta E_{rot} = I_s\ \Omega\ \Delta\Omega
\end{equation}
Combining these equations, the dependence on $I_p$ cancels, giving,
\begin{equation}
\Delta E_{rot} = -\left[mr_0^2 \Omega^2 \right] \frac{\Delta t}{3T}
\end{equation}
For both HS0705+6700 and HW~Vir, the average timing residuals of the substellar companion model fits to the observed O-C data are both of order 10 sec over 10 yr ($\Delta t/3T\sim 10^{-8}$), while the bracketed term evaluates to $3\cdot10^{40}$ J, so the change in rotational energy is 
\begin{equation}
\Delta E_{rot} \approx\ -3\cdot10^{32}\ \rm{J}
\label{eqn:E_loss}
\end{equation}
We can compare this with the maximum energy available from luminosity changes over the timescale for O-C variability in the secondary star. From Table~\ref{tab:Binary_props}, 
\begin{equation}
E_{max} = L\cdot t \sim 0.002\ L_{sun}\cdot 10\rm{\ yr} \sim 3\cdot 10^{32}\ J
\end{equation}
Remarkably, this is the same as the energy loss required to account for residual changes in the O-C model fits, but would be only $\sim10\%$ of the entire O-C deviations from a constant period were due to an Applegate-Lanza mechanism. This suggests that the observed O-C variations could be explained by a combination of substellar companions accounting for the bulk of the ETV's, with a Lanza-Applegate mechanism  contributing about 10\% of the total.

\subsection{Tidal Synchronization Timescale}
Another relevant dynamical timescale is the tidal synchronization time i.e., the timescale in which tides on the secondary dissipate the kinetic energy of asynchronous rotation \citep{Lanza:2020},
\begin{equation}
t_{sync} = 
\frac{2Q}{9} 
\frac{M_T}{m_p^2}
\frac{I_s}{\sqrt{GM_TR}}
{ \left( \frac{a}{R} \right) } ^{9/2}
\end{equation}
where Q is a dimensionless tidal quality factor that characterizes the efficiency of tidal energy dissipation in the active (secondary) component, $M_T$ is the total mass of the binary system, $m_p$ is the primary mass, $a$ is the binary semi-major axis, and $R$ is the active star radius.  Following \cite{Lanza:2020}, we assume $Q\sim10^5-10^6$  corresponding to strong tidal coupling due to nearly synchronous rotation. For the  moment of inertia $I_s$, we use the numerical expression of \cite{Criss:2015} for a fully convective secondary star with a polytropic index n = 1.5,

\begin{equation}
    I_s = 0.2\ m_{sec}R_{sec}^2
\end{equation}

Using stellar parameters list in Table~\ref{tab:Binary_props}, we find tidal synchronization timescales $t_s$ = 4 -- 40 yr and 10 -- 100 yr for HS0705 and HW~Vir respectively. These times are comparable to the observed O-C variations, so tidal synchronization may dictate the timescale for spin-orbit coupling rather than magnetic activity timescales.
\section{Summary and Conclusions}
In this paper, we have reported new eclipse timing observations of three post-common envelope binary systems. We combined these observations with previously published timing data to investigate whether the eclipse timing variations from a linear ephemeris could be explained by the gravitational influence of orbiting substellar companions. We used models comprising multiple substellar companions in coplanar orbits parallel to the observer’s line of sight, adjusting their orbital parameters and masses to best fit the observed timing variations. 

For one of the three binaries (HS2231+2441), nearly all times of minima are consistent with a linear ephemeris. Hence, there is no evidence for substellar companions in this system, although they cannot be ruled out if the masses are small. However for the remaining two binaries, there are large timing variations (ETV) from the linear trend, indicating the presence of one or more gravitationally perturbing effects.  We focused on finding orbital parameters and masses of companions that could account for the observed ETV’s by varying the masses and orbital elements to best-fit the observed ETV time signature using a least-squares minimization.

In order to ensure that the substellar components had stable orbits, the minimization function included a Hill stability ‘penalty’ ansatz that increases sharply when the minimum separation between adjacent components was smaller than one Hill radius. We also used a Bayesian inference algorithm to better characterize the uncertainties of the derived orbital parameters and masses. In addition, we used the Hill stability criterion $H>1$ as a prior to restrict orbital solutions only to those that were plausibly long-term stable i.e., at least as long as the  assumed time scale for common-envelope evolution \citep{Schreiber:2003, Ivanova:2013}. 

Finally, we checked stability timescales for each model using a numerical N-body symplectic time-step integrator (WHFAST) and an algorithm to detect the onset of chaotic behavior (MEGNO). We ran both the numerical integrations and chaos indicator time evolutions for $10^7$ years. 

For HS0705+6700, we found one and two-component models that reasonably reproduced the observed ETV time history and were stable on long time scales ($\approx 10^7$ yr). However, the fits were not statistically convincing, with reduced chi-squares of the fitted models in the range 10--19, although this may be partly a result of underestimated uncertainties in previously published timing measurements \citep[cf.][]{Hinse:2014}. 

For HW~Vir, we found no stable solutions that could fit the ETV time signature, although it was possible to find adequate four component fits, albeit with highly unstable orbits. This has also been the conclusion of several previous studies of HW~Vir \citep[e.g.,][]{Horner:2012, Brown-Sevilla:2021} and although not definitive, this strongly suggests that substellar companions cannot account for all of the ETV time history observed in these (and likely other) PCEB binaries. 

Therefore, we investigated whether some fraction of the observed variations could be caused by an Applegate mechanism in which the secondary star's gravitational quadrupole moment is altered by magnetic activity. 
Although the original Applegate mechanism was found to be energetically insufficient to account for the observed timing variations in sdB binaries (e.g., \citealt{Volschow:2016}), various revisions of the mechanism have been suggested that may account for at least a fraction of the observed variations. 

We applied the model of \cite{Lanza:2020}, in which  spin–orbit coupling is produced by an active (secondary) star's non-axisymmetric internal magnetic field creates a non-axisymmetric gravitational quadrupole moment. Using this model, we find that the predicted timing variations could account for perhaps 10\%  of the observed O-C variability, although this result is uncertain, since it is dependent on poorly constrained binary system parameters such as the tidal energy dissipation and the internal mass distribution of the stars. Hence, we posit that the observed O-C variations could result from a hybrid model in which substellar companions account for a largest fraction of the timing variations, but that an Appleget-Lanza mechanism contributes the remainder.  

\section*{Acknowledgements}

This research made use of the LMFIT nonlinear least-square minimization Python library, and Astropy, a community developed core Python package for Astronomy. We also made use of \texttt{emcee}, an MIT licensed pure-Python implementation of the  affine invariant Markov chain Monte Carlo (MCMC) ensemble sampler \citep{Goodman:2010}. We also used the excellent Python module \texttt{corner.py} to visualize multidimensional results of  Markov Chain Monte Carlo simulations \citep{Foreman-Mackey:2016}. Numerical orbit integrations in this paper made use of the REBOUND N-body code \citep{Rein:2012}. The simulations were integrated using WHFAST, a Wisdom-Holman symplectic integrator \citep{Rein:2015a}. 

We are grateful to David Pulley, John Mallett and the Altair Group for their contribution of times of minima for HS0705+6700 on Dec 29th, 2019, and times of minima for HW Vir on June 6th, 2019. We are also grateful to Prof. Antonino Lanza for several very insightful communications. The Iowa Robotic Observatory is funded by the College of Liberal Arts at the University of Iowa.

\section*{Data Availability}

We have provided Table \ref{tab:Minimum_times} as our new mid-eclipse timing data of three post-common envelope binaries. The light curves available on request from the corresponding author.




\bibliographystyle{mnras}
\bibliography{Binaries-library} 

\clearpage




\appendix

\label{Appendix}
\section{Time of minimum Uncertainty estimates}
In this appendix, we evaluate the uncertainty in the time of  minimum determination of an eclipsing binary light curve constructed from a series of images with relatively long exposure times and image download times. These pertain especially to observations using relatively small-aperture telescopes, including those reported in this paper. 

A standard technique used by many previous authors (more than 1,100 citation!) is the folded-symmetry algorithm described by \cite{Kwee:1956}. This technique determines the time of minimum by adjusting a trial midpoint time and minimizing the summed squared magnitude difference between points equally time-displaced on either side of the midpoint. The associated uncertainty is determined by fitting a quadratic function and applying standard propagation of errors to fitted minimum value. The technique relies on equally-space data samples and does not account for interpolation errors when this is not the case. Other more sophisticated methods \cite[e.g.,][]{Carter:2008,Deeg:2015,Deeg:2020} have addressed eclipse timing uncertainty, but they generally have been restricted to trapezoidal light curves, as observed in exoplanet transits. This is not applicable to short-period binary systems, whose eclipse light curves are much closer to Gaussian profiles \citep[e.g.,][]{Baran:2018}.

\begin{figure*}
\centering
\includegraphics[angle=0,width=0.95\textwidth]{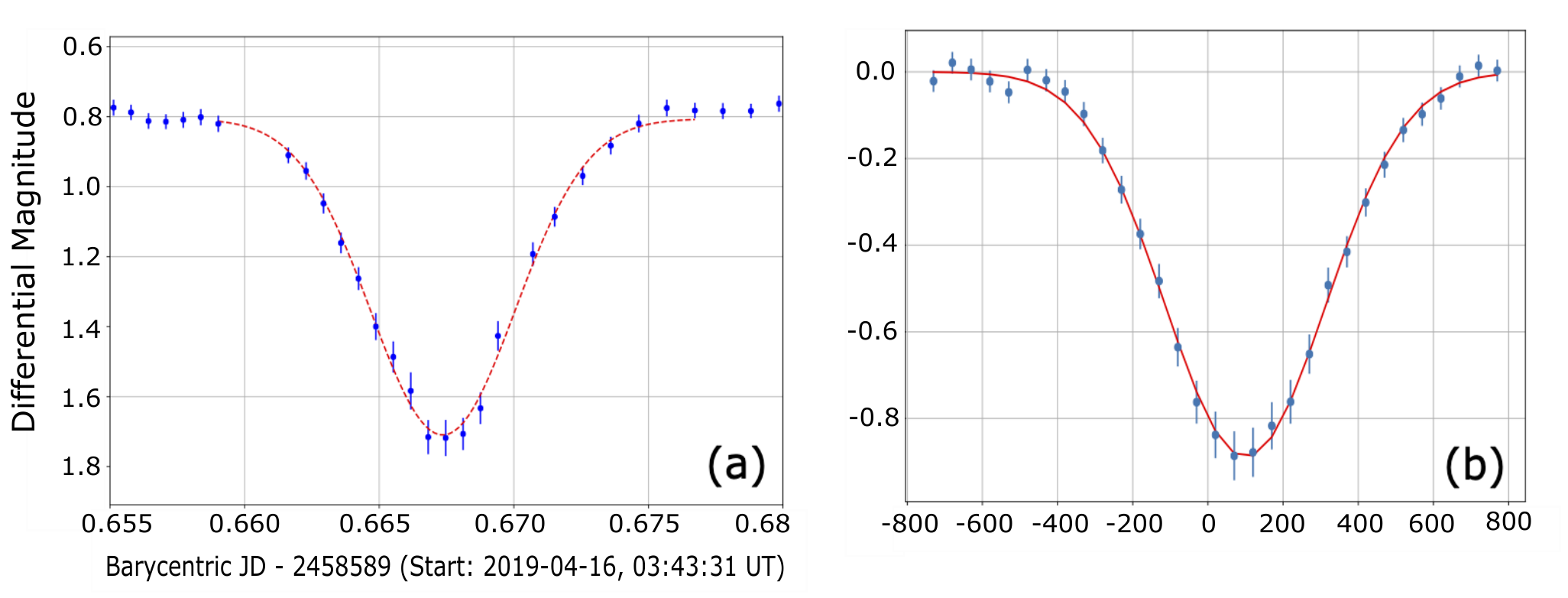}
\caption{(a) HS0705+6700 primary eclipse 16 April 2019 with LSQ fit Gaussian model.(b) Simulated light curve.}
\label{fig:lc-real-simulated}
\end{figure*}

We evaluated time of minimum uncertainty using a Monte Carlo approach, simulating observed eclipse light curves with a Gaussian profile whose width, depth, and scatter closely resembles the observed light curve. Fig.~\ref{fig:lc-real-simulated} shows a light curve of HS0705+6700 observed on with the Iowa Robotic telescope, along with the corresponding simulated light curve. The synthetic light curve was made by sampling a Gaussian function with 0.25 magnitude eclipse depth, and 300 sec half-width. We added Gaussian random noise to each sampled point, with a standard deviation $\sigma = 0.025\times 2.5^{\Delta m}$ where $\Delta m$ is the differential magnitude difference to the out-of-eclipse level. The synthetic light curve was sampled every 10 msec. We then binned the simulated date into 40 second bins with 10 second time gap between bins, corresponding the the observed image exposure time and download times for HS0705+6700. The magnitude assigned to each image was the computed mean value of 10 msec samples within each bin. The resulting synthetic light curve was fit with a Gaussian profile to determine the time of minimum. This process was repeated 10,00 times with different times of minima varying from -100 sec to +100 sec relative to the nominal zero-point time of minimum. The difference between the true time of minimum and the fitted time of minimum was recorded. 

\begin{figure*}
    \centering
    \includegraphics[width=0.75\textwidth]{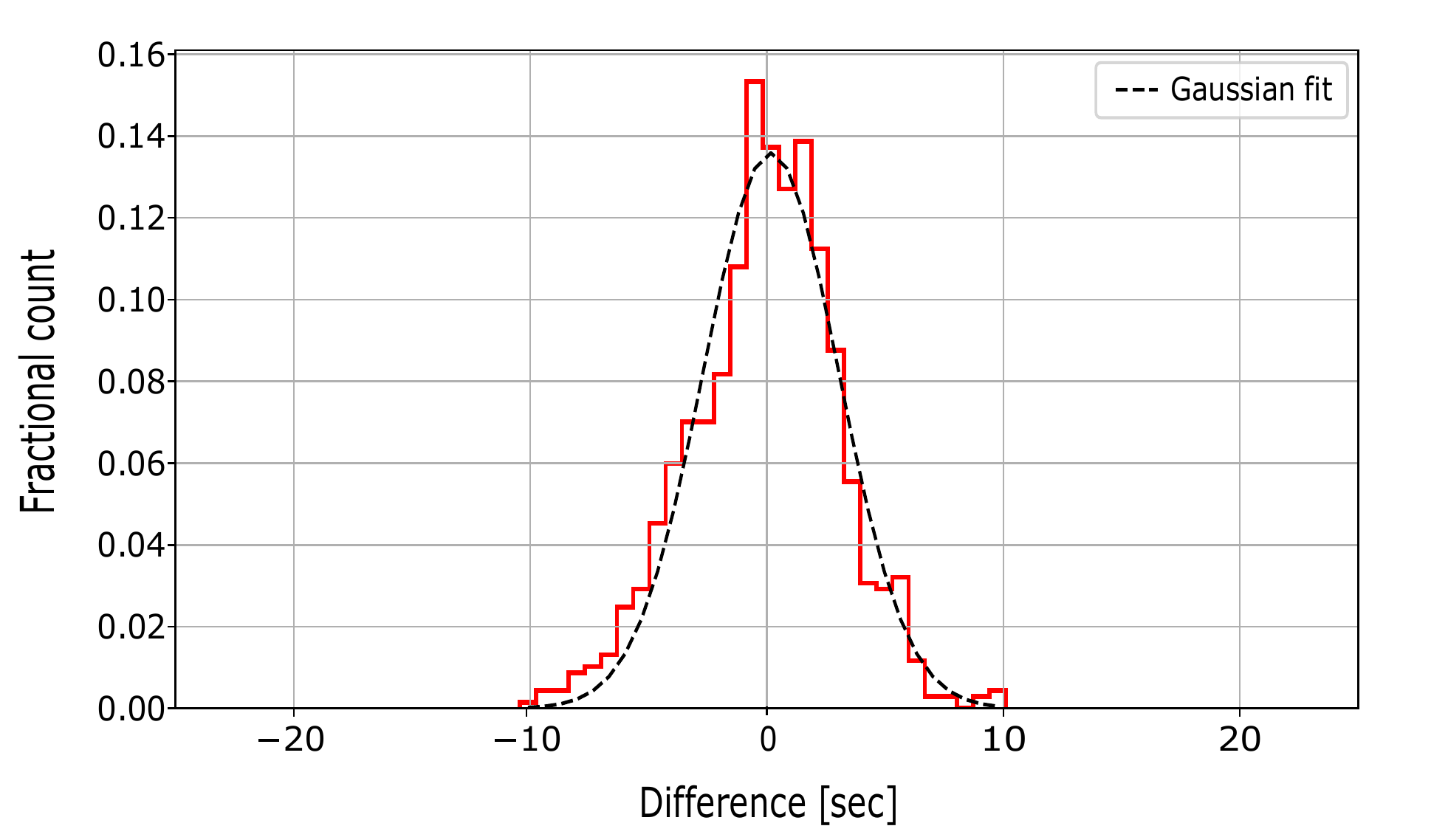}
    \caption{Histogram of time difference between true time of minimum and 10,000 Monte Carlo simulations.}
    \label{fig:histogram}
\end{figure*}

A histogram of the time difference between the best-fit time of minimum and the true minimum for 1,000 realizations is shown in Fig.\ref{fig:histogram}. The $1\sigma$ width is 4 seconds.

\bsp	
\label{lastpage}
\end{document}